\documentclass[10pt,conference]{IEEEtran}
\usepackage{amsmath}
\usepackage{amsfonts}
\usepackage{amssymb}
\usepackage{psfrag}
\usepackage{cite}

\newtheorem{theorem}{Theorem}
\newtheorem{lemma}{Lemma}

\newtheorem{assumption}{Assumption}
\newtheorem{problem}{Problem}
\newtheorem{property}{Property}

\def\QED{\hfill $\blacksquare$}
\begin{document}
 \newcommand{\paperstatus}{Submitted to \textit{IEEE Transactions on Signal Processing }}
\title{Full Large-Scale Diversity Space Codes for MIMO Optical Wireless Communications}

\author{
\authorblockN{Yan-Yu Zhang\IEEEauthorrefmark{1}, Hong-Yi Yu\IEEEauthorrefmark{1}, Jian-Kang Zhang\IEEEauthorrefmark{2}, Yi-Jun Zhu\IEEEauthorrefmark{1}, Jin-Long Wang\IEEEauthorrefmark{3} and Tao Wang\IEEEauthorrefmark{1}}
\authorblockA{\IEEEauthorrefmark{1}
Zhengzhou Information Science and Technology Institute, Zhengzhou, Henan, China\\
Emails: yyzhang.xinda@gmail.com; maxyucn@sohu.com; yijunzhu1976@gmail.com and yjswangtao@163.com}
\authorblockA{
\IEEEauthorrefmark{2}McMaster University, Hamilton, ONT L8S 4K1, Canada\\
Email: jkzhang@mail.ece.mcmaster.ca}
\authorblockA{
\IEEEauthorrefmark{3}University of Science and Technology, Nanjing, Jiangsu, China\\
Email: wjl543@sina.com}}

\maketitle
\begin{abstract}
In this paper, we consider a multiple-input-multiple-output optical wireless communication (MIMO-OWC) system suffering from log-normal fading. In this scenario, a general criterion for the design of full large-scale diversity  space code (FLDSC) with the maximum likelihood (ML) detector is developed. Based on our criterion, FLDSC is attained if and only if all the entries of the space coding matrix are positive. Particularly for $2\times 2$ MIMO-OWC with  unipolar pulse amplitude modulation (PAM), a closed-form linear FLDSC satisfying this criterion is attained by smartly taking advantage of some available properties as well as by developing some new interesting properties on Farey sequences in number theory to rigorously attack the continuous and discrete variables mixed max-min problem. In fact, this specific design not only proves that a repetition code (RC) is the best linear FLDSC, but also uncovers a significant difference between MIMO radio frequency (RF) communications and  MIMO-OWC that space-only transmission is sufficient for a full diversity achievement. Computer simulations demonstrate that  FLDSC substantially outperforms  spatial multiplexing with the same total optical power and spectral efficiency and the latter   obtains  only the small-scale diversity gain.
\end{abstract}
\section{Introduction}
Optical wireless communications (OWC), due to its potential for bandwidth-hungry applications, has become a very important area of research~\cite{o2005optical,chan2006free,das2008requirements,Kumar2010Led-based,Elgala2011review,Borah2012review, Gancarz13}. However, some challenges remain, especially in atmospheric environments, where \textit{robustness} is a key consideration. Therefore, in the design of high date rate OWC links, we need to consider the atmospheric impairments-induced fading which can be described by the log-normal (LN) statistical model ~\cite{Beaulieu2008itct,giggenbach2008fading}. To combat fading, multi-input-multi-output (MIMO) OWC (MIMO-OWC) systems introduce the design for the transmitted symbols distributed over transmitting apertures (space) and (or) symbol periods (time). Full large-scale diversity is achieved when the total degrees of freedom (DoF) available in the MIMO-OWC system is fully utilized.

Unfortunately, unlike MIMO techniques for  radio frequency (MIMO-RF) communications with Rayleigh fading, there are \textit{two} significant challenges in MIMO-OWC communications. The \textit{first} is that there does not exist any available mathematical tool that could be directly applied to the analysis of the average pair-wise error probability (PEP) when LN is involved. Although there are really  mathematical formulae  in literature for numerically and accurately computing the integral involving LN~\cite{haas2002space,navidpour2007itwc,Beaulieu2008itct}, it can not be used for the theoretic analysis on diversity. The \textit{second} is a \textit{nonnegative constraint} on the design of transmission for MIMO-OWC, which is a major difference between MIMO RF communications and MIMO-OWC. It is because of this constraint that the currently available well-developed MIMO techniques for RF communications can not be directly utilized for MIMO-OWC.  Despite the fact that the nonnegative constraint can be satisfied by properly adding some direct-current components (DC) into transmitter designs so that the existing advanced MIMO techniques~\cite{tarokh98} for RF communications such as orthogonal space-time block code (OSTBC)~\cite{alamouti98,tarokh99} could be used in MIMO-OWC, the power loss arising from DC incurs the fact that these modified OSTBCs~\cite{simon2005alamouti,wang2009mimo} in a LN fading optical channel have worse error performance than the RC~\cite{navidpour2007itwc,majid2008twc,abaza2014diversity}.

All the aforementioned factors greatly motivate us to develop a general criterion on the design of full large-scale diversity transmission for MIMO-OWC. As an initial exploration, we consider the  space-alone code,  and intend  to uncover some unique characteristics of MIMO-OWC by establishing a general criterion for the design of FLDSC and attaining an optimal analytical solution to a specific two by two linear FLDSC.

\section{Channel Model And Space Code}\label{sec:model}
\subsection{Channel model with space code}
Let us consider an $M\times N$ MIMO-OWC system having $M$ receiver apertures and $N$ transmitter apertures transmitting the symbol vector $\mathbf{s}$, $\{s_l\},l=1,\ldots,L$, which are randomly, independently and equally likely, selected from a given constellation. To facilitate the transmission of these $L$ symbols through the $N$ transmitters in the one time slots (channel use), each symbol $s_l$ is mapped by a space encoder $\mathbf{F}_{l}$ to an $N\times 1$  space code vector $\mathbf{F}\left(s_l\right)$ and then summed together, resulting in an $N\times 1$ space codeword given by $\mathbf{x}=\sum_{l=1}^L\mathbf{F}_l\left(s_l\right)$, where the $n$-th element of $\mathbf{x}$ represents the coded symbol to be transmitted from the $n$-th transmitter aperture. These coded symbols are then transmitted to the receivers through flat-fading path coefficients, which form the elements of the $M\times N$ channel matrix $\mathbf{H}$. The received space-only symbol, denoted by the $M\times 1$ vector $\mathbf{y}$, can be written as
 \begin{eqnarray}\label{eqn:system_model}
\mathbf{y}=\frac{1}{P_{op}}\mathbf{H}\mathbf{x}+\mathbf{n},
\end{eqnarray}
  where $P_{op}$ is the average optical power of $\mathbf{x}$ and,  the entries of  channel matrix $\mathbf{H}$ are independent and LN distributed, i.e., $h_{ij}=e^{z_{ij}}$, where $z_{ij}\sim\mathcal{N}\left(\mu_{ij},\sigma_{ij}^2\right), i=1\ldots M,j=1\ldots N$.
  The probability density function (PDF) of $h_{i,j}$ is
\begin{eqnarray}
f_{H}\left(h_{ij}\right)=\frac{1}{\sqrt{2\pi}h_{ij}\sigma_{ij} }\exp\left(-\frac{\left(\ln h_{ij}-\mu_{ij}\right)^2}{2\sigma _{ij}^{2}}\right)
\end{eqnarray}
The PDF of $\mathbf{H}$ is $f_{\mathbf{H}}\left(\mathbf{H}\right)=\prod_{i=1}^{M}\prod_{j=1}^{N}f_{H}\left(h_{ij}\right)$.
  The signalling scheme of $\mathbf{s}$ is unipolar pulse amplitude modulation (PAM) to meet the unipolarity requirement of intensity modulator (IM), i.e., $\mathbf{x}\in\mathbb{R}_+^{N\times 1}$. As an example, the constellation of unipolar $2^p$-ary PAM is $\mathcal{B}_{2^p}=\{0,1,\ldots,2^p-1\}$, where $p$ is a positive integer. Then, the equivalent constellation of $\mathbf{s}$ is $\mathcal{S}=\{\mathbf{s}:s_i\in \mathcal{B},i=1,\ldots N\}$, i.e., ${\mathcal S}={\mathcal B}_{2^p}^N$.

Furthermore, for noise vector $\mathbf{n}$, the two primary sources at the receiver front end are due to noise from the receive electronics and shot noise from the received DC photocurrent induced by background  radiation~\cite{Karp1988,Barry1994Ifrd}. By the central limit theorem, this high-intensity shot noise for the lightwave-based OWC is closely approximated as additive, signal-independent, white, Gaussian noise (AWGN)~\cite{Barry1994Ifrd} with zero mean and variance $\sigma _{n}^{2}$.

 By rewriting the channel matrix  as a vector and aligning the code-channel product to form a new channel vector, we can have $\mathbf{Hx}=\left(\mathbf{I}_{M}\otimes\mathbf{x}^T\right)\textrm{vec}\left(\mathbf{H}\right)
$, where $\otimes$ denotes the Kronecker product operation and $\textrm{vec}\left(\mathbf{H}\right)=\left[h_{11},\ldots,h_{1N},\ldots,h_{M1},\ldots,h_{MN}\right]^T$.
 For discussion convenience, we call $\mathbf{I}_{M}\otimes\mathbf{x}^T$ a codeword matrix, denoted by $\mathbf{ S}\left(\mathbf{x}\right)$. Then, the correlation matrix of the corresponding error coding matrix is given by
   \begin{eqnarray}\label{eqn:rank_one_equivalence}
           \mathbf{ S}^T\left(\mathbf{e}\right)\mathbf{ S}\left(\mathbf{e}\right)=
\mathbf{I}_{M}\otimes\mathbf{X}\left(\mathbf{e}\right)
   \end{eqnarray}
   where $\mathbf{X}\left(\mathbf{e}\right)=\mathbf{e}\mathbf{e}^T$, $\mathbf{e}=\mathbf{F}\left(\mathbf{\hat{s}}\right)-\mathbf{F}\left(\mathbf{s}\right)$ is the error vector with $\mathbf{s}\neq\mathbf{\hat{s}}$ and $\mathbf{s},\mathbf{\hat{s}}\in\mathcal{S}$. All these non-zero $\mathbf{e}$ form an error set, denoted by $\mathcal{E}$.

   \subsection{Problem formulation}
To formally state our problem, we make the following assumptions throughout this paper.
\begin{enumerate}
\item \textit{Power constraint}. The average optical power is constrained, i.e., $E\left[\sum_{i}^{N}x_{i}\right]=P_{op}$. Although limits are placed on both the average and peak optical power transmitted, in the case of most practical modulated optical sources, it is the average optical power constraint that dominates~\cite{hranilovic2003optical}.
\item \textit{SNR  definition}. The optical SNR is defined by $\rho_{op}=\frac{P_{op}}{\sqrt{N\sigma_n^2}}$, since the noise variance per dimension is assumed to be $\sigma_n^2$. Thus, in expressions on error performance involved in the squared Euclidean distance, the term $\rho$, in fact, is equal to
     \begin{eqnarray}\label{eqn:electrical_snr}
       \rho=\frac{1}{N\sigma_n^2}
     \end{eqnarray}
     with optical power being normalized by $\frac{1}{P_{op}}$. Unless stated otherwise, $\rho$ is referred to as  the squared optical SNR thereafter.
\end{enumerate}

 Under the above  assumptions, our primary task in this paper is to establish a general criterion on the design of FLDSC and  solve the following problem.
\begin{problem}\label{prob:design_problem} Design the space encoder $\mathbf{F}(\cdot)$ subject to the total optical power such that 1) $\forall \mathbf{s}\in \mathcal{S}, \mathbf{F}\left(\mathbf{s}\right) $ meets the unipolarity requirement of IM; 2) Full large-scale diversity is enabled for the ML receiver.~\hfill\QED
\end{problem}

\section{Design Criteria for Space Code}

This subsection aims at deriving the PEP of MIMO-OWC  and then, establishing a general design criterion for the linear space coded system.
\subsection{PEP of MIMO-OWC}\label{sec:performance_analysis}
 Given a channel realization $\mathbf{H}\in\mathbb{R}_{+}^{M\times N}$ and a transmitted signal vector $\mathbf{s}$, the probability
   of transmitting $\mathbf{s}$ and deciding in favor of $\hat{\mathbf s}$ with the ML receiver is given by~\cite{forney98}
\begin{eqnarray}\label{eqn:ml_detection_pep1}
P\left(\mathbf{s}\rightarrow\mathbf{\hat{s}}|\mathbf{H}\right)=Q\left(\frac{d\left(\mathbf{e}\right)}{2}\right) \end{eqnarray}
where $d^2\left(\mathbf{e}\right)=\frac{\rho}{NP_{op}^2}\textrm{vec}\left(\mathbf{H}\right)^T\mathbf{ S}^T\left(\mathbf{e}\right)\mathbf{ S}\left(\mathbf{e}\right)\textrm{vec}\left(\mathbf{H}\right)=\frac{\rho}{NP_{op}^2}\sum_{i=1}^M\left(\mathbf{h}_i^T\mathbf{e}\right)^2$ with $\mathbf{h}_i=\left[h_{i1},\ldots,h_{iN}\right]^T,i=1,\ldots,M$. Averaging \eqref{eqn:ml_detection_pep1}  over $\mathbf{H}$ yields
\begin{eqnarray} \label{eqn:ml_detection_pep2}
P\left(\mathbf{s}\rightarrow\mathbf{\hat{s}}\right)
&=&\int P\left(\mathbf{s}\rightarrow\mathbf{\hat{s}}|\mathbf{H}\right)f_{\mathbf{H}}\left(\mathbf{H}\right)d\mathbf{H}.
\end{eqnarray}

 To extract the dominant term of~\eqref{eqn:ml_detection_pep2},  we make an assumption for time being. Later on, we will prove that this condition is actually necessary and sufficient for $\mathbf{ X}\left(\mathbf{e}\right)$ to render full diversity.

\begin{assumption} \label{assumpt:existence_of_rectangular}
Any $\mathbf{e}\in {\mathcal E}$ is unipolar without zero entry.~\hfill\QED
\end{assumption}

\begin{theorem}\label{theorem:pep_mimo_owc}
 Under Assumption \ref{assumpt:existence_of_rectangular},
$P\left(\mathbf{s}\rightarrow\mathbf{\hat{s}}\right)$ is bounded by
  \begin{eqnarray}\label{eqn:pep_mimoowc}
  &&\underbrace{C_{L} \left(\ln\rho\right)^{-MN}e^{-\sum_{i=1}^{M}\sum_{j=1}^{N}\frac{\left(\ln\rho +\ln \left(P_{op}^2\Omega\right)-\ln\left(M\sum_{k=1}^Ne_k^2\right)\right)^2}{8\sigma_{ij}^2}}}_{P_{L}\left(\mathbf{s}\rightarrow\mathbf{\hat{s}}\right)}
\nonumber\\
    &&\le P\left(\mathbf{s}\rightarrow\mathbf{\hat{s}}\right) \le \underbrace{C_{U1}
\rho^{-\frac{MN}{2}}
e^{-\sum_{i=1}^{M}\sum_{j=1}^{N}\frac{\ln^2 \rho}{8\sigma_{ij}^2  }}}_{P_{U1}\left(\mathbf{s}\rightarrow\mathbf{\hat{s}}\right)}
\nonumber\\
&&+\underbrace{ C_{U2}\left(\ln\rho\right)^{-MN}e^{-\sum_{i=1}^{M}\sum_{j=1}^{N}\frac{\left(\ln \frac{\rho}{\ln^2 \rho} +\ln \left(P_{op}^2\Omega\right)-\ln e_j^2\right)^2}{8\sigma_{ij}^2}}}_{P_{U2}\left(\mathbf{s}\rightarrow\mathbf{\hat{s}}\right)}
\end{eqnarray}
where $\Omega=\sum_{i=1}^{M}\sum_{j=1}^{N}\sigma_{ij}^{-2}$,
$C_{L}=\frac{\prod_{i=1}^M\prod_{j=1}^N\sigma_{ij}}{\left(4\pi\right)^{MN}e^{-\frac{MN}{2}}}Q\left(\frac{1}{2}\left(\sum_{k=1}^Ne_k^2\right)^{-\frac{1}{2}}\right)$, $C_{U1}=\frac{e^{\frac{\sum_{i=1}^M\sum_{j=1}^N\sigma_{ij}^2}{2}}}{2\prod_{i=1}^N\prod_{j=1}^M\sigma_{ij}}
\left(\frac{\sum_{k=1}^{N}e_k^2}{NP_{op}^2}\right)^{-\frac{MN}{2}}$ and $C_{U2}=\frac{\left(NP_{op}^2\right)^{MN}}{2\prod_{i=1}^M\prod_{j=1}^N\sqrt{\sigma_{ij}^2}}e^{-\frac{\Omega}{8}\ln^2\left(\frac{NP_{op}^2\Omega}{M}\right)}$.
~\hfill\QED
\end{theorem}

Now, we can see that in \eqref{eqn:pep_mimoowc}, $P_{L}\left(\mathbf{s}\rightarrow\mathbf{\hat{s}}\right)$ and $P_{U2}\left(\mathbf{s}\rightarrow\mathbf{\hat{s}}\right)$ have the same exponential term, $\exp\left(-\frac{\Omega}{8}\ln^2 \rho\right)$, whereas the exponential term of $P_{U1}\left(\mathbf{s}\rightarrow\mathbf{\hat{s}}\right)$ is $\exp\left(-\frac{\Omega}{8}\ln^2\frac{\rho}{\ln^2\rho}\right)$, which decays slower than  $\exp\left(-\frac{\Omega}{8}\ln^2 \rho\right)$ against high SNR. That being said, we have successfully attained the dominant term, $P_{U1}\left(\mathbf{s}\rightarrow\mathbf{\hat{s}}\right)$, of the upper-bound of $P\left(\mathbf{s}\rightarrow\mathbf{\hat{s}}\right)$. Thus, our selection of $\tau$ is reasonable to capture the dominant behaviour of $P\left(\mathbf{s}\rightarrow\mathbf{\hat{s}}\right)$.

With all the aforementioned preparations, we enable to give the general design criterion for FLDSC of MIMO-OWC in the following subsection.
 \subsection{Design Criterion for FLDSC}

  The discussions in Subsection~\ref{sec:performance_analysis} tells us that $P_{U1}\left(\mathbf{s}\rightarrow\mathbf{\hat{s}}\right)$  is the dominant term of the upper-bound of $P\left(\mathbf{s}\rightarrow\mathbf{\hat{s}}\right)$ in \eqref{eqn:pep_mimoowc}. With this, we will provide a guideline on the space code design in this subsection.  To define the performance parameters to be optimized, we rewrite $P_{U2}\left(\mathbf{s}\rightarrow\mathbf{\hat{s}}\right)$ as follows.
\begin{eqnarray}\label{eqn:dominant_term}
P_{U2}\left(\mathbf{s}\rightarrow\mathbf{\hat{s}}\right)=
C_{U2}\mathcal{G}_{c}\left(\mathbf{e}\right)
\left(\frac{\rho}{\ln^2 \rho}\right)^{\frac{\Omega}{4}\ln\left(
\frac{NP_{op}^2\Omega}{M}\right)-\frac{3}{4}\ln \mathcal{G}_{d}\left(\mathbf{e}\right)}&&\nonumber\\
\times
\left(\ln \rho\right)^{-MN}\exp\left(-\frac{\Omega}{8} \ln^2 \frac{\rho}{\ln^2 \rho}\right)&&
\end{eqnarray}
where $\mathcal{G}_{d}\left(\mathbf{e}\right)=\prod_{j=1}^{N}|e_j|^{\sum_{i=1}^M\sigma_{ij}^{-2}}$ and $\mathcal{G}_{c}\left(\mathbf{e}\right)=\exp\left( \frac{1}{2}\sum_{i=1}^{M}\sum_{j=1}^{N}
\left(\ln |e_j|^{\sigma_{ij}}\right)^2\right)\left(\frac{NP_{op}^2\Omega}{M}\right)^{
\frac{1}{2}\ln\ln \mathcal{G}_{d}\left(\mathbf{e}\right)}$.

Here, the following three factors dictate the minimization of $P_{U1}\left(\mathbf{s}\rightarrow\mathbf{\hat{s}}\right)$:
\begin{enumerate}
  \item \textit{Large-scale diversity gain}. The exponent $\Omega$ with respect to $\ln \frac{\rho}{\ln^2 \rho}$ governs the behavior of $P_{U1}\left(\mathbf{s}\rightarrow\mathbf{\hat{s}}\right)$. For this reason,  $\Omega$ is named as the \textit{large-scale diversity gain}. The full large-scale diversity achievement is equivalent to the event that all the $MN$ terms in $\Omega=\sum_{i=1}^{M}\sum_{j=1}^{N}\sigma_{ij}^{-2}$ offered by the $N\times M$ MIMO-OWC are fully utilized.  Thus, when we design space code, full large-scale diversity must be assured \textit{in the first place}.
 \item \textit{Small-scale diversity gain}. $\mathcal{G}_{d}\left(\mathbf{e}\right)=\prod_{j=1}^{N}|e_j|^{\sum_{i=1}^M\sigma_{ij}^{-2}}$ is called \textit{small-scale diversity gain}, which affects the polynomial decaying in terms of $\frac{\rho}{\ln^2 \rho}$. $\min_{\mathbf{e}}\mathcal{G}_{d}\left(\mathbf{e}\right)$ should be maximized to optimize the error performance of the worst error event. Since the small-scale diversity gain will affect the average PEP via the polynomially decaying speed of the error curve, the small-scale diversity gain of the space code  is what to be optimized \textit{in the second place}.
 \item \textit{Coding gain.} $\mathcal{G}_{c}\left(\mathbf{e}\right)$ is defined as\textit{ coding gain}. On condition that both diversity gain are maximized, if there still exists DoF for further optimization of the coding gain, $\max_{\mathbf{e}\in\mathcal{E}}\mathcal{G}_{c}\left(\mathbf{e}\right)$ should be minimized as the \textit{last step} for the systematical design of space code.
\end{enumerate}

In what follows, we will give a sufficient and necessary condition on a full large-scale diversity achievement. Hence, Assumption \ref{assumpt:existence_of_rectangular} is \textit{sufficient and necessary} for FLDSC, which is summarized as the following theorem:
 \begin{theorem} \label{theorem:space_code_full_diversity}
A space code enables full large-scale diversity if and only $\forall \mathbf{e}\in \mathcal{E}$, $\mathbf{e}$ is unipolar without zero-valued entries or equivalently, $\forall \mathbf{e}\in \mathcal{E}$, $\mathbf{ X}\left(\mathbf{e}\right)$ is positive.~\hfill\QED
 \end{theorem}

With these results, we can proceed to design FLDSC systematically in the following section.

\section{Optimal Design of Specific Linear FLDSC }\label{sec:design_example}
In this section, we will exemplify our established criterion in~\eqref{eqn:dominant_term} by designing  a specific \textit{linear} FLDSC for $2\times 2$ MIMO-OWC with  unipolar pulse amplitude modulation (PAM). For this particular design, a closed-form space code optimizing both diversity gains will be obtained by smartly taking advantage of some available properties as well as by developing some new interesting properties on Farey sequences in number theory.

\subsection{Design Problem Formulation}
 Consider a $2\times2$ MIMO-OWC system with $\mathbf{F}\left(\mathbf{s}\right)=\mathbf{F}\mathbf{s}$, where
$  \mathbf{F} =
  \left({
  \begin{array}{cc}
  f_{11}&  f_{12}\\
  f_{21}&  f_{22}\\
  \end{array}
  }\right)$ and $\mathbf{ X}\left(\mathbf{e}\right)=\left(
{\begin{array}{cc}
e_{1}^2&e_1e_2\\
e_1e_2&e_2^2\\
\end{array}
}\right)$.
By Theorem~\ref{theorem:space_code_full_diversity}, $\mathbf{ X}\left(\mathbf{e}\right)$ should be positive to maximize the large-scale diversity gain.
On the other hand, from  the structure of $\mathbf{ X}\left(\mathbf{e}\right)$ and \eqref{eqn:dominant_term},  the small-scale diversity gain is $\mathcal{G}_{d}\left(\mathbf{e}\right)=|e_1e_2|$ under the assumption that CSIT is unknown.
Therefore, to optimize the worst case over $\mathcal{E}$, FLDSC design is  formulated as follows:
    \begin{eqnarray}\label{eqn:modulator_design}
&&\max_{f_{11},f_{12},f_{21},f_{22}} \min_{\mathbf{e}} e_1e_2\nonumber \\
&& s.t.
\left\{
  \begin{array}{ll}
\left[e_1,e_2\right]^T\in \mathcal{E},f_{ij}>0,i,j\in\{1,2\},\\
e_1e_2>0,f_{11}+f_{12}+f_{21}+f_{22}=1.
  \end{array}
\right.
\end{eqnarray}

Our task is to  analytically  solve \eqref{eqn:modulator_design}.
To do that, we first simplify \eqref{eqn:modulator_design}  by finding all the possible minimum terms.

\subsection{Equivalent  Simplification of Design Problem}\label{subsec:simplification}
For $2^{p}$-PAM, all the possible non-zero values of $e_1e_2$ are
\begin{eqnarray}\label{eqn:objective_function}
e_1e_2=\left(mf_{11}\pm nf_{12}\right)\left(mf_{21}\pm nf_{22}\right)\neq0,m,n\in\mathcal{B}_{2^p}.
\end{eqnarray}
\subsubsection{Preliminary simplification}
After observations over  \eqref{eqn:objective_function},  we have the following  facts.
 \begin{enumerate}
   \item  $\forall m\neq0,m,n\in\mathcal{B}_{2^p}$, it holds holds that
\begin{subequations}
   \begin{eqnarray}
\left(mf_{11}+ nf_{12}\right)\left(mf_{21}+nf_{22}\right)
  \ge f_{11}f_{21}.
   \end{eqnarray}
   \item $\forall n\neq0$, $m,n\in\mathcal{B}_{2^p}$, it is true that
   \begin{eqnarray}
\left(mf_{11}+ nf_{12}\right)\left(mf_{21}+nf_{22}\right)\ge f_{12}f_{22}.
   \end{eqnarray}
   \item $\forall k\neq0,m^2+n^2\neq0,k,m,n\in\mathcal{B}_{2^p}$, we have
   \begin{eqnarray}
\frac{k\left(mf_{11}-  nf_{12}\right)\left( mf_{21}- nf_{22}\right)}{\left(mf_{11}-nf_{12}\right)\left(mf_{21}-nf_{22}\right)}
   \ge 1.
   \end{eqnarray}
   \end{subequations}
 \end{enumerate}
So,  all the possible minimum of $e_1e_2$ in  \eqref{eqn:modulator_design}  are
$f_{11}f_{21}$, $f_{12}f_{22}$ and $\left(mf_{11}-nf_{12}\right)\left(mf_{21}-nf_{22}\right)$,
where $\frac{n}{m}$ are irreducible, i.e., $m\perp n$. These terms are denoted by
$F_{10}=f_{12}f_{22}\left(\frac{f_{11}}{f_{12}}\times\frac{f_{21}}{f_{22}}\right),F_{01}=f_{12}f_{22}
$ and $F_{mn}=f_{12}f_{22}\left(m\frac{f_{11}}{f_{12}}-n\right)\left(m\frac{f_{21}}{f_{22}}-n\right)$.
After putting aside the common term, $f_{12}f_{22}$, we can see that $F_{mn}$ is the piecewise linear function of $\frac {f_{11}}{f_{12}}$ and $ \frac{f_{21}}{f_{22}}$, respectively. So,  \eqref{eqn:modulator_design} can be solved by fragmenting interval $\left[0,\infty\right)$ into disjoint subintervals. This fragmentation can be done by  the breakpoints where $F_{mn}=0$.
To  characterize this sequence, there exists an elegant mathematical tool in number theory presented below.
   \subsubsection{Farey sequences}
First, we observe some specific examples of the breakpoint sequences.
For OOK, the breakpoints $\frac{0}{1}, \frac{1}{1},\infty$.
For 4-PAM, they are $\frac{0}{1},\frac{1}{3},\frac{1}{2},\frac{2}{3},\frac{1}{1},\frac{3}{2},\frac{2}{1},\frac{3}{1},\infty$.
 For  8-PAM, we have the breakpoint sequence with the former part being
   \begin{subequations}
   \begin{eqnarray}\label{eqn:before_1}
\frac{0}{1},\frac{1}{7},\frac{1}{6},\frac{1}{5},\frac{1}{4},
\frac{2}{7},\frac{1}{3},\frac{2}{5},\frac{3}{7},\frac{1}{2},
\frac{4}{7},\frac{3}{5},\frac{2}{3},\frac{5}{7},
  \frac{3}{4},\frac{4}{5},\frac{5}{6},\frac{6}{7},\frac{1}{1}&&
  \end{eqnarray}
and the remaining being
  \begin{eqnarray}\label{eqn:after_1}
 \frac{7}{6},\frac{6}{5},\frac{5}{4},\frac{4}{3}, \frac{7}{5},\frac{3}{2},\frac{5}{3},\frac{7}{4},\frac{2}{1},
  \frac{7}{3},\frac{5}{2},\frac{3}{1},\frac{7}{2},\frac{4}{1},\frac{5}{1},\frac{6}{1}
              ,\frac{7}{1},\infty&&
   \end{eqnarray}
   \end{subequations}
  Through  these special examples, we find that the series of breakpoints before $1/ 1$ (such as the sequence in \eqref{eqn:before_1}) is
the one which is called the Farey sequence~\cite{hardy1979introduction}.
The Farey sequence $\mathfrak{F}_k$ for any positive integer $k$ is the set
 of irreducible rational numbers $\frac{a}{b}$ with $0\leq a\leq b\leq k$ arranged in an increasing order.
 The series of breakpoints after $\frac{1}{1}$ (such as the sequence in \eqref{eqn:after_1}) is
the reciprocal version of the Farey sequence. Thus, our  focus is on the sequence before $\frac{1}{1}$.

The Farey sequence has many interesting properties~\cite{hardy1979introduction}, some of which closely relevant to our problem are given as follows.
\begin{lemma}\label{lemma:farey_sequence}
If $\frac{n_1}{ m_1}$, $\frac{n_2}{ m_2}$ and $\frac{n_3}{ m_3}$ are three successive terms of $\mathfrak{F}_k,k>3$ and $\frac{n_1}{ m_1}<\frac{ n_2}{ m_2}<\frac{ n_3}{ m_3}$,
then,
\begin{enumerate}
  \item $ m_1n_2-m_2n_1=1$ and $m_1+m_2\ge k+1$.
  \item $\frac{n_1+n_2}{m_1+m_2}\in\left(\frac{n_1}{m_1},\frac{n_3}{m_3}\right)$ and $\frac{n_2}{m_2}=\frac{n_1+n_3}{m_1+m_3}$.
\end{enumerate}
~\hfill\QED
\end{lemma}

However, having only Lemma \ref{lemma:farey_sequence} is not enough to solve our  design problem in \eqref{eqn:modulator_design}. We need to develop the other new properties of Farey sequences, concluded by Properties~\ref{property:local_two_worst_cases}, \ref{property:the_local_solution} and \ref{th:min}.

\begin{property} \label{property:local_two_worst_cases} Given $k>3$, assume $\frac{n_0}{ m_0},\frac{n_1}{m_1},\frac{n_2}{ m_2},\frac{n_3}{ m_3}\in\mathfrak{F}_{k}$ and $\frac{n_0}{m_0}<\frac{n_1}{m_1}<\frac{n_2}{m_2}<\frac{n_3}{m_3}$.
If $\frac{n_1}{m_1}$ and $\frac{n_2}{ m_2}$  are successive,
then, $\frac{n_1+n_3}{m_1+m_3}\ge\frac{n_2}{m_2}$ and $\frac{n_0+n_2}{m_0+m_2}\le\frac{n_1}{m_1}$.
~\hfill\QED
\end{property}
\begin{property} \label{property:the_local_solution}
Assume $\frac{n_1}{m_1},\frac{n_2}{m_2}\in \mathfrak{F}_{k}, k>3$ and $\frac{n_1}{m_1}<\frac{n_2}{m_2}$.   Then,
  \begin{enumerate}
     \item $\frac{n_1}{m_1}<\frac{n_1+n_2}{m_1+m_2}<\frac{n_2}{m_2}$ holds.
      \item If $\frac{f_{11}}{f_{12}},\frac{f_{21}}{f_{22}}\in\left(\frac{n_1}{m_1},\frac{n_1+n_2}{m_1+m_2} \right)$, then,  $F_{m_1n_1}<F_{m_2n_2}$.
      \item If $\frac{f_{11}}{f_{12}},\frac{f_{21}}{f_{22}}\in\left(\frac{n_1+n_2}{m_1+m_2},\frac{n_2}{m_2} \right)$, then,  $F_{m_1n_1}>F_{m_2n_2}$.
      \item If $\frac{f_{11}}{f_{12}}=\frac{f_{21}}{f_{22}}=\frac{n_1+n_2}{m_1+m_2}$, then, $F_{m_1n_1}=F_{m_2n_2}$.
    \end{enumerate}
~\hfill\QED
\end{property}

Using Properties~\ref{property:local_two_worst_cases} and~\ref{property:the_local_solution}, we attain the following property.
\begin{property}\label{th:min}
If $\frac{n_1}{m_1}$ and $\frac{n_2}{m_2}$ are successive in $\mathfrak{F}_{k}$ and $\frac{f_{11}}{f_{12}},\frac{f_{21}}{f_{22}}\in\left(\frac{n_1}{m_1},\frac{n_2}{m_2} \right)$,
then, $  F_{m_1n_1}$ and $F_{m_2n_2}$ are the two worst cases.~\hfill\QED
\end{property}

\subsection{Techniques to Solve The Max-min Problem}\label{subsec:max_min}
  Thanks to Farey sequences, \eqref{eqn:modulator_design} is transformed into a piecewise max-min problem  with two objective functions. By solving this kind of problem, our code construction results can be presented as the following theorem.
\begin{theorem}\label{theorem:golbal_solution}
The solution  to~\eqref{eqn:modulator_design} is determined by
\begin{eqnarray}\label{eqn:global_optimal_modulator}
  \mathbf{F} =\frac{1}{2+2^{p+1}}\left(
  {\begin{array}{ccc}
  1&2^p\\
  1&2^p\\
 \end{array}}
  \right),
   \textrm{or}~\frac{1}{2+2^{p+1}}\left(
  {\begin{array}{ccc}
  2^p&1\\
  2^p&1\\
 \end{array}}
  \right).
\end{eqnarray}
~\hfill \QED
\end{theorem}

Theorem~\ref{theorem:golbal_solution} uncovers the fact that the optimal linear space coded symbols are actually unipolar $2^{2p}$-ary PAM symbols,
since $\mathcal{B}_{2^{2p}}=\{s_1+2^p s_2:s_1,s_2\in\mathcal{B}_{2^p}\}$. Therefore, in fact, we have rigorously proved that RC~\cite{navidpour2007itwc} is optimal in the sense of the criterion established in this paper.

\section{Computer Simulations }\label{sec:numerical_results}

In this section, we carry out computer simulations to verify our newly developed criterion in \eqref{eqn:dominant_term}. In light of our work being initiative, the only space-only transmission scheme available in the literature is spatial multiplexing (SM). Accordingly,  we compare the performance of spatial multiplexing and FLDSC specifically designed for $2\times2$ MIMO-OWC in Section \ref{sec:design_example}.
In addition, we suppose that $h_{ij},i,j=1,2$ are independently and identically distributed and let $\sigma_{11}=\sigma_{12}=\sigma_{21}=\sigma_{22}=\sigma$.
 These schemes are  as follows:
 \begin{enumerate}
  \item \textit{FLDSC}.  The optical power is  normalized in such a way that $\sum_{i,j=1}^2f_{ij}=2$ yields $E\left[\sum_{i,j=1}^2 f_{ij}s_j\right]=1$. From  \eqref{eqn:global_optimal_modulator}, the coding matrix is
$  \mathbf{F} =
  \frac{1}{3}\left({
  \begin{array}{cc}
  2&  1\\
  2&  1\\
  \end{array}
  }\right)$.
  \item \textit{SM}.  We fix the modulation formats to be OOK and vary $\sigma^2$. So the rate is 2 bits per channel use (pcu). The  transmitted symbols $s_1,s_2$ are chosen from $\{0,1\}$ equally likely. The average optical power is $E\left[s_1+s_2\right]=1$.
  \end{enumerate}
\begin{figure}[!htp]
    \centering
    \resizebox{7cm}{!}{\includegraphics{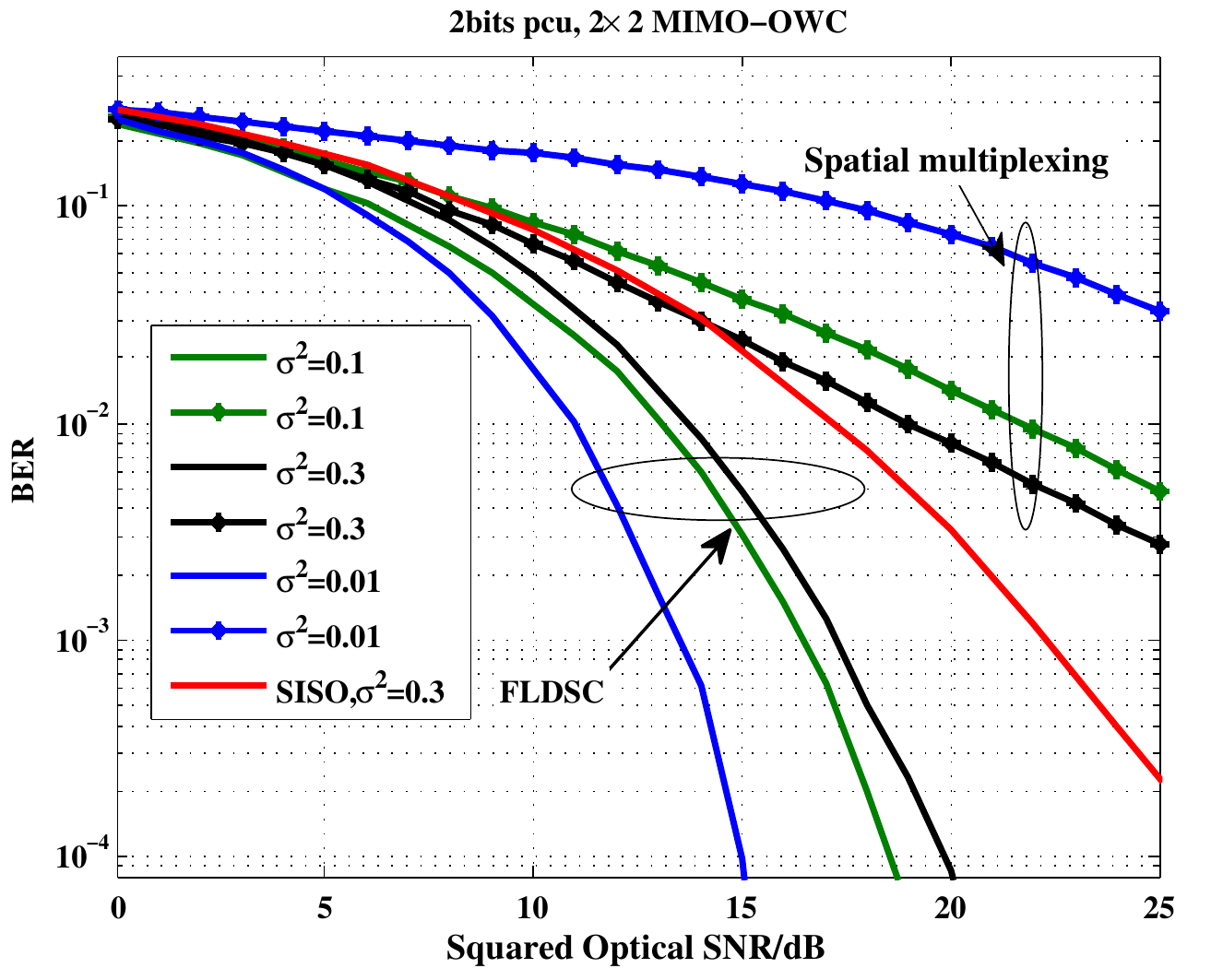}}
    \centering \caption{BER  comparisons of  FLDSC and spatial multiplexing.}
    \label{fig:modulated_unmodulated}
\end{figure}
\begin{figure}[!htp]
    \centering
    \resizebox{7cm}{!}{\includegraphics{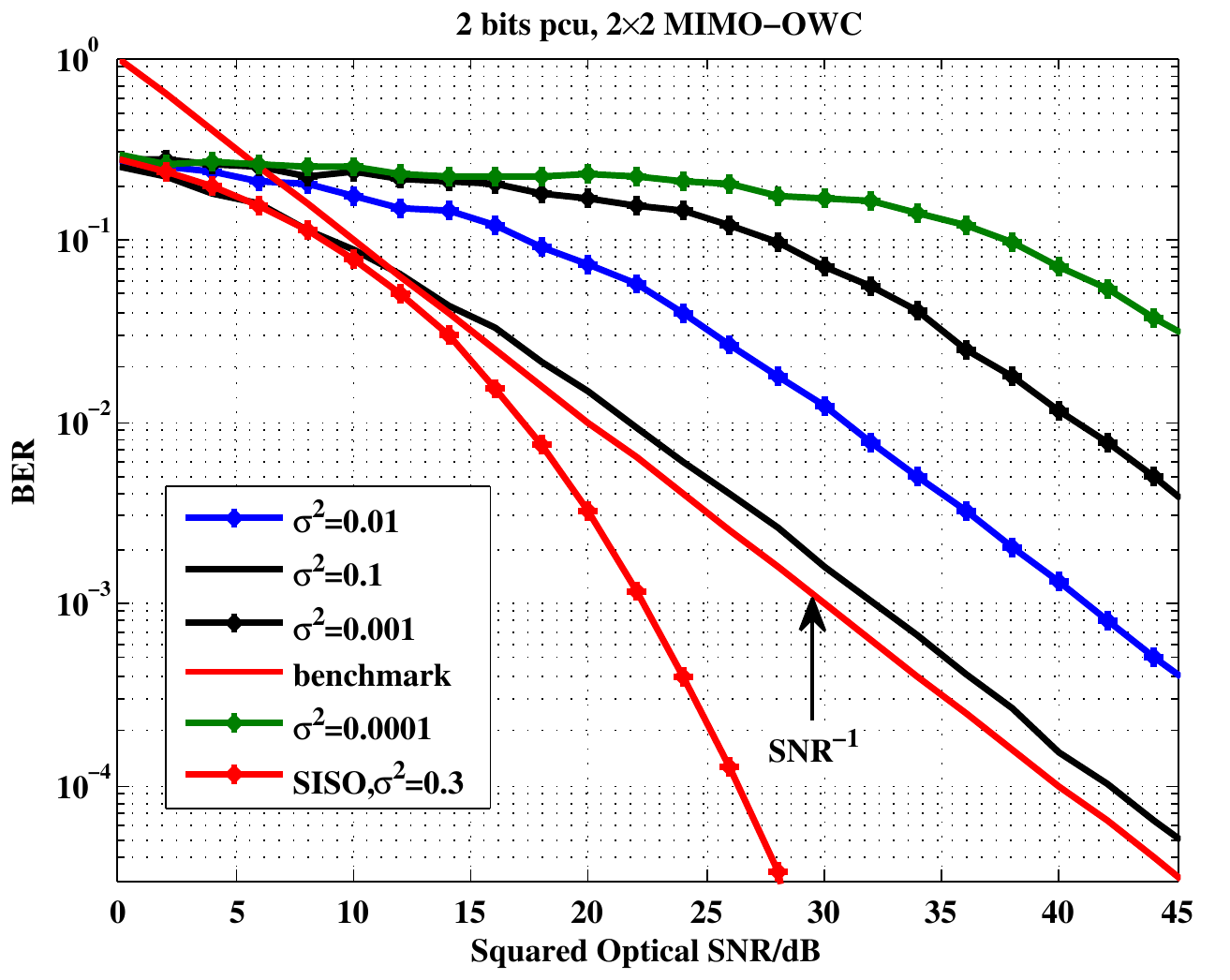}}
    \centering \caption{BER performance  of spatial multiplexing.}
    \label{fig:multiplexing_MIMO}
\end{figure}
 We can see that both schemes  have the same spectrum efficiency, i.e., 2 bits pcu and the same optical power. Through numerical results, we have following observations.

 Substantial enhancement from FLDSC is achieved, as shown in Fig. \ref{fig:modulated_unmodulated}. For $\sigma^2=0.01$, the improvement is almost 16 dB at the target bit error rate (BER) of $10^{-2}$. For $\sigma^2=0.5$, the improvement is almost 6 dB at the target BER of $10^{-3}$. Note that the small-scale gain also governs the negative slope of error curve. The decaying speed of the error curve of FLDSC is exponential in terms of $\ln\frac{\rho}{\ln^2\rho}$, whereas that of SM is polynomial with respect to $\rho$, even worse than single-input-singal-output (SISO).

 SM presents only small-scale diversity gain illustrated in Fig. \ref{fig:multiplexing_MIMO}.  By varying the variance of $\mathbf{H}$, we find that in the high SNR regimes, the error curve decays as $\rho^{-1}$ as long as the SNR is high enough . From $\sigma^2=0.001$ to $\sigma^2=0.1$, the error curve has a horizonal shift, which is the typical style of RF MIMO~\cite{tarokh98}. The reason is given as follows. The equivalent space coding matrix is
      $\mathbf{ X}\left(\mathbf{e}\right)=\left(
{\begin{array}{cc}
e_{1}^2&e_1e_2\\
e_1e_2&e_2^2\\
\end{array}
}\right),e_1,e_2\in\{0,\pm1\}$ with $e_1^2+e_2^2\neq0$. It should be noted that there exists two typical error events: $e_1e_2=-1$ and $e_1e_2=0$. From the necessity proof of Theorem \ref{theorem:space_code_full_diversity}, for $e_1e_2=-1$, the attained large-scale diversity gain is zero, and at the same time, if $e_1e_2=0$ with $e_1^2+e_2^2\neq0$, then, the attained large-scale diversity gain is only two for $2\times2$  MIMO-OWC. Therefore, the overall large-scale diversity gain of SM is zero with small-scale diversity gain being attained.

\section{Conclusion and Discussions}
In this paper, we have established a general criterion on the full-diversity  space coded transmission of MIMO-OWC for the ML receiver, which is, to our best knowledge, the first design criterion for the full large-scale diversity transmission of optical wireless communications with IM/DD over log-normal fading channels. Particularly for a $2\times 2$ case, we have attained an optimal closed-form FLDSC, rigorously proving that RC is the best among all the linear space codes. Our results clearly indicate that the transmission design is indeed necessary and essential for significantly improving the overall error performance for MIMO-OWC.

\section{Acknowledgements}
This work was supported in part by   Key Laboratory of Universal Wireless  Communications (Beijing University of Posts and Telecommunications), Ministry of Education of P. R. China under Grant No. KFKT-2012103, in part by NNSF of China (Grant No. 61271253) and in part by NHTRDP of China (``863'' Program) (Grant No. 2013AA013603).
\bibliographystyle{ieeetr}

\bibliography{tzzt}

\begin{thebibliography}{10}

\bibitem{o2005optical}
D.~O'Brien and M.~Katz, ``Optical wireless communications within
  fourth-generation wireless systems [invited],'' {\em Journal of optical
  networking}, vol.~4, no.~6, pp.~312--322, 2005.

\bibitem{chan2006free}
V.~W. Chan, ``Free-space optical communications,'' {\em J. Lightw. Technol.},
  vol.~24, no.~12, pp.~4750--4762, 2006.

\bibitem{das2008requirements}
S.~Das, H.~Henniger, B.~Epple, C.~I. Moore, W.~Rabinovich, R.~Sova, and
  D.~Young, ``Requirements and challenges for tactical free-space lasercomm,''
  in {\em Proc. IEEE Milit. Commun. Conf.}, pp.~1--10, 2008.

\bibitem{Kumar2010Led-based}
N.~Kumar and N.~R. Lourenco, ``Led-based visible light communication system: a
  brief survey and investigation,'' {\em J. Eng. Appl. Sci}, vol.~5,
  pp.~297--307, 2010.

\bibitem{Elgala2011review}
H.~Elgala, R.~Mesleh, and H.~Haas, ``Indoor optical wireless
  communication:potential and state-of-the-art,'' {\em IEEE Commun. Mag.},
  pp.~56--62, Sep. 2011.

\bibitem{Borah2012review}
D.~K. Borah, A.~C. Boucouvalas, C.~C. Davis, S.~Hranilovic, and K.Yiannopoulos,
  ``A review of communication-oriented optical wireless systems,'' {\em EURASIP
  J. Wireless Commun. Netw.}, vol.~91, pp.~1--28, 2012.

\bibitem{Gancarz13}
J.~Gancarz, H.~Elgala, and T.~D.~C. Little, ``Impact of lighting requirements
  on {VLC} systems,'' {\em IEEE Commun. Mag.}, pp.~34--41, Dec. 2013.

\bibitem{Beaulieu2008itct}
N.~C. Beaulieu and Q.~Xie, ``An optimal lognormal approximation to lognormal
  sum distributions,'' {\em IEEE Trans. Commun. Technol.}, vol.~53,
  pp.~479--489, 2004.

\bibitem{giggenbach2008fading}
D.~Giggenbach and H.~Henniger, ``Fading-loss assessment in atmospheric
  free-space optical communication links with on-off keying,'' {\em Opt.
  Engineering}, vol.~47, no.~4, pp.~046001--046001, 2008.

\bibitem{haas2002space}
S.~M. Haas, J.~H. Shapiro, and V.~Tarokh, ``Space-time codes for wireless
  optical communications,'' {\em EURASIP J. Appl. Signal Process.}, vol.~2002,
  no.~1, pp.~211--220, 2002.

\bibitem{navidpour2007itwc}
S.~M. Navidpour, M.~Uysal, and M.~Kavehrad, ``{BER} performance of free-space
  optical transmission with spatial diversity,'' {\em IEEE Trans. Wireless
  Commun.}, vol.~6, pp.~2813--2819, Aug. 2007.

\bibitem{tarokh98}
V.~Tarokh, N.~Seshadri, and A.~R. Calderbank, ``Space-time codes for high date
  rate wireless communication: performance criterion and code construction,''
  {\em IEEE Trans. Inf. Theory}, vol.~44, pp.~744--765, Mar. 1998.

\bibitem{alamouti98}
S.~M. Alamouti, ``A simple transmit diversity scheme for wireless
  commincations,'' {\em IEEE J. Select. Areas Commun}, vol.~16, pp.~1451--1458,
  Oct. 1998.

\bibitem{tarokh99}
V.~Tarokh, H.~Jafarkhani, and A.~R. Calderbank, ``Space-time block codes from
  orthogonal designs,'' {\em IEEE Trans. Inf. Theory}, vol.~45, pp.~1456--1467,
  July. 1999.

\bibitem{simon2005alamouti}
M.~K. Simon and V.~A. Vilnrotter, ``Alamouti-type space-time coding for
  free-space optical communication with direct detection,'' {\em IEEE Trans.
  Wireless Commun.}, vol.~4, no.~1, pp.~35--39, 2005.

\bibitem{wang2009mimo}
H.~Wang, X.~Ke, and L.~Zhao, ``{MIMO} free space optical communication based on
  orthogonal space time block code,'' {\em Science in China Series F:
  Information Sciences}, vol.~52, no.~8, pp.~1483--1490, 2009.

\bibitem{majid2008twc}
M.~Safari and M.~Uysal, ``Do we really need {OSTBC}s for free-space optical
  communication with direct detection?,'' {\em IEEE Trans. Wireless Commun.},
  vol.~7, pp.~4445--4448, November 2008.

\bibitem{abaza2014diversity}
M.~Abaza, R.~Mesleh, A.~Mansour, and E.-H.~M. Aggoune, ``Diversity techniques
  for a free-space optical communication system in correlated log-normal
  channels,'' {\em Opt. Engineering}, vol.~53, no.~1, 2014.

\bibitem{Karp1988}
S.~Karp, R.~M. Gagliardi, S.~E. Moran, and L.~B. Stotts, ``Optical channels,''
  {\em Opt. Engineering}, vol.~47, no.~4, pp.~046001--046001, 2008.

\bibitem{Barry1994Ifrd}
J.~R. Barry, {\em Wireless Infrared Communications}.
\newblock Boston, MA: Kluwer Academic Press, 1994.

\bibitem{hranilovic2003optical}
S.~Hranilovic and F.~R. Kschischang, ``Optical intensity-modulated direct
  detection channels: signal space and lattice codes,'' {\em IEEE Trans. Inf.
  Theory}, vol.~49, no.~6, pp.~1385--1399, 2003.

\bibitem{forney98}
G.~D. Forney and G.~U. Ungerboeck, ``Modulation and coding for linear
  {G}aussian channel,'' {\em IEEE Trans. Inf. Theory}, vol.~44, pp.~2384--2415,
  May 1998.

\bibitem{hardy1979introduction}
G.~H. Hardy and E.~M. Wright, {\em An introduction to the theory of numbers}.
\newblock Oxford University Press, 1979.

\end{thebibliography}
\small

\end{document}